\documentclass[preprintnumbers,groupedaddress,nofootinbib,aps]{revtex4}
\usepackage[latin1]{inputenc}
\usepackage{graphicx}
\usepackage{color}
\usepackage{multirow}
\usepackage{amsmath,amssymb}
\usepackage{hyperref}
\usepackage{url}
\usepackage{slashed}
\usepackage{subfigure}
\usepackage[usenames,dvipsnames]{xcolor}
\newcommand{\be}{\begin{equation}}
\newcommand{\ee}{\end{equation}}
\newcommand{\bea}{\begin{eqnarray}}
\newcommand{\eea}{\end{eqnarray}}
\newcommand{\tf}{\tilde{f}}
\newcommand{\tR}{\tilde{R}}
\newcommand{\gb}{\bar{g}}
\newcommand{\Db}{\bar{D}}
\newcommand{\Rb}{\bar{R}}
\newcommand{\Eqref}[1]{Eq.~\eqref{#1}}
\begin{document}
\title{The Renormalization Group flow of unimodular f(R) gravity}

\author{Astrid Eichhorn}
\affiliation{\mbox{\it Blackett Laboratory, Imperial College, Prince Consort Road, London SW7 2AZ, United Kingdom}
\mbox{\it E-mail: {a.eichhorn@imperial.ac.uk}}}

\begin{abstract}
Unimodular gravity is classically equivalent to General Relativity. This equivalence extends to actions which are functions of the curvature scalar. At the quantum level, the dynamics could differ. Most importantly, the cosmological constant is not a coupling in the unimodular action, providing a new vantage point from which to address the cosmological constant fine-tuning problem. Here, a quantum theory based on the asymptotic safety scenario is studied, and evidence for an interacting fixed point in unimodular $f(R)$ gravity is found.
We study the fixed point and its properties, and also discuss the compatibility of unimodular asymptotic safety with dynamical matter, finding evidence for its compatibility with the matter degrees of freedom of the Standard Model.\end{abstract}

\maketitle

\section{Introduction}
There are only very few observations from which one could expect to learn something about the deep structure of spacetime, described by a model of quantum gravity.
 One of those is the observed accelerated expansion of the universe, which can be modelled by including a nonzero cosmological constant in the Einstein-Hilbert action. This entails the cosmological constant problem. Here, we will focus on one aspect of this problem, namely the question why  quantum vacuum fluctuations do not seem to gravity, i.e., why the cosmological constant exhibits a severe fine-tuning problem. As its mass-dimensionality is 2, one would expect quantum fluctuations to drive it to be of order one in units defined by the square of the physical mass scale of the theory, which is the Planck scale. In terms of Renormalization Group (RG) trajectories, the tiny value inferred from observations implies that a particularly fine-tuned trajectory has to be picked. Of course every so-called relevant coupling in a quantum theory corresponds to one free parameter that can only be fixed by comparison with experiment. Thus one always has to pick a particular RG trajectory in order for the model to reproduce observations. On the other hand, relevant couplings with only, e.g., a logarithmic running imply that if one picks a trajectory nearby, the measured value of the coupling will only change slightly. For the cosmological constant, this statement is not true, i.e., for a reason which as yet has no dynamical explanation, our universe just "happens" to live on a highly fine tuned choice of trajectory.

As already proposed by Weinberg \cite{Weinberg:1988cp}, a "degravitation" of the cosmological constant is possible by changing its status from a coupling in the action to a constant of integration that arises at the level of the equations of motion.  The second is a "classical" quantity in the sense that it is not affected by quantum fluctuations. In unimodular gravity \cite{Einstein}, the metric is conceived as a symmetric tensor with fixed determinant $\sqrt{-g} = \epsilon$ \cite{Unruh:1988in}. This implies that no operator of the form $\sqrt{-g}$ exists, as the volume is just a fixed number, and the cosmological constant is removed from the space of couplings, the theory space. Once a Renormalization Group trajectory in this reduced space has been picked, the effective equations of motion can be calculated from the full effective action -- the infrared endpoint of the trajectory -- and the cosmological constant will then make its appearence as a constant of integration. In this way, quantum vacuum fluctuations do not affect the value of the cosmological constant.
It is thus of interest to investigate a quantum theory of unimodular gravity. Unimodular gravity in both its quantum and classical form has sparked considerable interest since it was originally proposed \cite{unimod_papers,Alvarez,vanderBij:1981ym,Unruh:1988in,Henneaux:1989zc,Smolin:2009ti,Smolin2011}.

A second motivation to consider unimodular quantum gravity lies in the fact it will most probably differ from the non-unimodular version of quantum gravity.
This inequivalence arises, as imposing the unimodularity condition alters the spectrum of fluctuations of the theory. In more detail, deriving the full metric propagator by taking the second variation of the action yields different results when $\sqrt{-g} = \epsilon$ is imposed, than if the metric determinant is allowed to fluctuate. In particular, fluctuations of the conformal mode, which yield an instability of the path-integral in the Euclidean case when starting from the Einstein-Hilbert action, are absent in unimodular gravity. This already suggests that although classically equivalent \cite{vanderBij:1981ym}, the quantum version of unimodular and ''standard" gravity could differ. The absence of the conformal instability even suggests that the unimodular quantum theory could have better properties.

As the search for the "quantum theory of gravity describing our universe" is still ongoing, an exploration of different models for quantum gravity models is clearly of interest, both from a theoretical as well as from a phenomenological point of view. Here we will focus on exploring models of asymptotically safe quantum gravity. As we will point out in Sec.~\ref{sec:mattermatters}, it might be possible to distinguish between different versions of asymptotically safe gravity experimentally.

Finally, in order to better understand the structure of Renormalization Group (RG) flows in gravity it is of interest to consider settings with fewer propagating degrees of freedom in the path-integral. Here, we should clarify that the physically propagating degrees of freedom in both settings, unimodular vs. "full" gravity, agree in a perturbative expansion around a flat background, i.e., there is a massless spin-2 graviton \cite{vanderBij:1981ym}. On the other hand, the configurations that enter the path integral are different metric configurations, and, e.g., the configuration space of the conformal mode is also summed over in the case of "full" gravity. In order to shed light on the physical mechanism of asymptotic safety, it is helpful to consider settings where some of the modes in the path-integral are removed. 

In this paper, we will consider unimodular asymptotic safety, and investigate truncated Renormalization Group flows based on an $f(R)$ action.
We will discuss the classical equivalence of "full" gravity, which we will call Einstein gravity, with unimodular gravity based on an $f(R)$ action in Sec.~\ref{sec:classicaleq}. We will then focus on the quantum theory, and review the asymptotic safety scenario and the functional RG in Sec.~\ref{sec:ASandFlow}, where we will also present all technical details of our calculation. In Sec.~\ref{sec:results} we will present the flow equation for $f(R)$ and discuss a fixed point and its properties. We will make a first step toward phenomenology in studying the effect of dynamical matter in Sec.~\ref{sec:mattermatters}, and finally conclude in Sec.~\ref{sec:conclusions}.

\section{Relation between unimodular gravity and Einstein gravity}
\subsection{Classical equivalence of $f(R)$ gravity and unimodular $f(R)$}\label{sec:classicaleq}
Before we embark on an analysis of quantum gravity, let us clarify the classical relation between unimodular $f(R)$ gravity and $f(R)$ gravity with a full metric, see, e.g., \cite{DeFelice:2010aj} for a review.
Here we will focus on the Lorentzian case, and then switch to a Euclidean setting for the analysis of the quantum theory. We focus on actions given by a  function of the curvature scalar, $f(R)$, with $f(0)=0$. We introduce the Newton coupling $G_N$ and the cosmological constant $\Lambda$ in the action
\be
S = \int d^4x \sqrt{-g} \left( f(R) + \frac{1}{8 \pi G_N}\Lambda + \mathcal{L}_m\right).
\ee
The corresponding equations of motion
are given by 
\be
- \frac{1}{2} f(R)g_{\mu \nu} + f'(R) R_{\mu \nu} - D_{\mu}D_{\nu}f'(R) + g_{\mu \nu}D^2 f'(R) + \frac{1}{16 \pi G_N} \Lambda g_{\mu \nu} = \frac{1}{2}\,T_{\mu \nu},\label{fulleom}
\ee
where the energy-momentum tensor is given by
\be
T_{\mu \nu}= -\frac{2}{\sqrt{-g}} \frac{\delta \mathcal{L}_m}{\delta g^{\mu \nu}}.
\ee
The Bianchi-identities
\be
D^{\mu} \left(R_{\mu \nu} - \frac{1}{2} g_{\mu \nu}R \right)=0,\label{Bianchi}
\ee
will now play a crucial role: By taking the covariant derivative of \Eqref{fulleom}, we deduce the conservation law for the energy-momentum tensor by imposing the Bianchi-identities.

To obtain the unimodular equations of motion for the action
\be
S_u = \int d^4x \epsilon \left(f(R) + \mathcal{L}_m \right),
\ee
we have to consider tracefree variations $g^{\mu \nu}\delta g_{\mu \nu}=0$ and obtain
\be
f'(R) R_{\mu \nu} - D_{\mu}D_{\nu} f'(R) + \frac{1}{4} g_{\mu \nu} D^2 f'(R) - \frac{1}{4} g_{\mu \nu}R f'(R) = \frac{1}{2} \left(T_{\mu \nu} - \frac{1}{4}g_{\mu \nu} T^{\lambda}_{\lambda} \right). \label{unimodeom}
\ee
Crucially, the covariant derivative of the lhs of \Eqref{unimodeom} does not vanish when the Bianchi-identities are used.
Instead, we can \emph{impose} conservation of the energy-momentum tensor\footnote{As discussed, e.g., in \cite{Alvarez}, it is not clear whether this requirement can be preserved in a quantum field theory setting for the matter degrees of freedom. This could potentially lead to a situation where the low-energy effective equations of motion allow us to distinguish between General Relativity and unimodular gravity.}, and thereby derive a nontrivial identity, namely
\be
D_{\nu} \left(-\frac{1}{4} f'(R)R - \frac{3}{4}D^2 f'(R) + \frac{1}{8} T^{\lambda}_{\lambda} \right) = \frac{-1}{2} D_{\nu}f(R),
\ee
where we have used \Eqref{unimodeom} and imposed \Eqref{Bianchi}.
This identity
 allows us to identify $1/2 f(R)$ with $-\frac{1}{4} f'(R)R - \frac{3}{4}D^2 f'(R) + \frac{1}{2} T^{\lambda}_{\lambda}$, up to a constant of integration, which we choose to call $\frac{1}{16 \pi G_N} \Lambda$. Inserting this identity into \Eqref{unimodeom} we obtain \Eqref{fulleom}, i.e., classically unimodular $f(R)$ gravity cannot be distinguished from standard $f(R)$ gravity. Note that this statement depends on the \emph{postulate} of energy-momentum conservation in the unimodular case. It is a priori clear that the two theories can only be classically equivalent, if one additional condition is imposed in the unimodular case: Since the equations of motion of unimodular gravity are obtained by removing the trace from the standard equations of motion, they contain precisely one condition less.

\section{Unimodular quantum gravity}\label{sec:ASandFlow}
\subsection{Asymptotic safety}\label{sec:AS}
From now on we will focus on Euclidean quantum gravity, as this allows for a straightforward application of RG tools. To arrive at a unimodular quantum theory of gravity, we will invoke the asymptotic safety conjecture for gravity \cite{Weinberg:1980gg}. 
Interestingly, unimodularity plays a role in a number of other approaches to quantum gravity, e.g., within causal set quantum gravity, where a discrete version of unimodularity would be implemented by performing the path-sum over all causets with a fixed number of elements \cite{Henson:2006kf,Wallden:2010sh}. Further, Causal Dynamical Triangulations is based on a setting where the number of simplices usually is held fixed for the simulations, i.e., the cosmological constant is removed from the space of couplings, see, e.g., \cite{Ambjorn:2014gsa}. 

An asymptotically safe quantum theory of gravity is valid at arbitrarily high momenta, i.e., beyond the regime of validity of effective field theory \cite{Donoghue:2012zc,Donoghue:1993eb}, and at the same time remains predictive, i.e., only comes with a finite number of free parameters. It is the second requirement that breaks down in a perturbative quantization of gravity when extended to arbitrarily high momenta \cite{'tHooft:1974bx,Goroff:1985sz,vandeVen:1991gw}. Then, an infinite number of counterterms needs to be introduced and there is no mechanism to predict the corresponding free parameters.
In asymptotic safety, the Renormalization Group flow approaches an interacting fixed point at high momenta. 
Specifically, we refer to the dimensionless couplings here, which can be obtained from the dimensionful ones by an appropriate rescaling with the RG momentum scale $k$. If these couplings approach a fixed point, the theory becomes scale free and can be extended to arbitrarily high momentum scales. 
The RG flow lives in an infinite dimensional space of all couplings, the theory space, as quantum fluctuations generically generate all operators that are compatible with the symmetries.  In this infinite-dimensional space one must then investigate whether predictivity can be obtained, i.e., whether the model has only a finite number of free parameters.
This is ensured, if the interacting fixed point comes with a finite number of relevant, i.e., ultraviolet (UV) attractive directions. 
The (ir)relevance of a coupling  $g_i(k)$ determines its scale-dependence 
in the vicinity of a fixed point. This can be obtained after linearising the RG flow:
\be
g_i(k)=g_{i\, \ast}+\sum_I C_I V_i^{I} \left(\frac{k}{k_0} \right)^{-\theta_I}.
\ee
Herein, $g_{i\, \ast}$ denotes the fixed-point values of the coupling $g_i$.
$V^I$ are the eigenvectors of the stability matrix $\mathcal{M}_{ij} =( \partial \beta_{g_i}/ \partial g_j) \vert_{g_n=g_{n\, \ast}}$, and $-\theta_I$ its eigenvalues. $C_I$ are constants of integration. If $\theta_J<0$, then $C_J=0$ is required in order for the couplings to approach the fixed point in the UV limit, where $k \rightarrow \infty$. On the other hand, relevant directions are those with $\theta_J>0$. These approach the fixed point automatically, imposing no requirement on the corresponding parameter $C_J$. They have instead to be determined experimentally.
 $k_0$ is an arbitrary reference scale, which can be taken as the scale at which the value of the couplings is measured.
At a Gau\ss{}ian fixed point, such as that underlying asymptotic freedom in Yang-Mills theory, only the couplings of positive and vanishing mass dimensionality can be relevant. At an interacting fixed point, quantum fluctuations shift the critical exponents away from the mass dimensionality by an anomalous dimension. Additionally, the (ir)relevant directions are typically no longer given by the operators defined at the Gau\ss{}ian fixed point, but in fact correspond to mixtures of these.
When the anomalous dimensions remain finite, this suggests that only a finite number of relevant directions exist, and that these can be found among appropriate combinations of the couplings with the largest mass dimensionalities. 

It remains for us to determine whether an interacting fixed point exists in unimodular gravity and which of the couplings correspond to relevant ones. Most importantly, we cannot draw any strong conclusions from the evidence for a fixed point in the case of quantum Einstein gravity \cite{Reuter:1996cp,AS_lit,Lauscher:2001ya,Codello:2007bd, 
Machado:2007ea,Benedetti:2012dx,Dietz:2012ic,Falls:2013bv,bimetric}. As the symmetry changes from full diffeomorphisms to transverse diffeomorphisms, and the field content is restricted by the unimodularity requirement, the two theory spaces are different. The existence and properties of possible fixed points can therefore be different in the two cases, as discussed in \cite{Eichhorn:2013xr}.

\subsection{Comments on the equivalence between quantum Einstein gravity and unimodular quantum gravity}
In the following, we will refer to the quantum theory with a full dynamical metric and full diffeomorphism symmetry as quantum Einstein gravity, also in cases where the underlying action is not the Einstein-Hilbert action.
In the unimodular case, $\sqrt{g} = \epsilon$ implies that the volume term is not an operator any more, but simply a number. As such, it will be dropped from the action, and no quantum fluctuations contribute to the running of the prefactor, i.e., the cosmological constant. In other words, the cosmological constant is removed from the unimodular theory space. 
In principle, one could also attempt to impose the unimodularity condition employing a Lagrange multiplier, in which case the cosmological constant would remain a running coupling, presumably resulting in an inequivalent quantum model \cite{Smolin:2009ti}.
In quantum Einstein gravity, one can impose $\sqrt{g} = \epsilon$ as a gauge. This does of course \emph{not} remove the cosmological constant from the theory space. This becomes particularly important when the cosmological constant corresponds to a relevant coupling at an RG fixed point. Then the predicted values of all other couplings in the infrared will depend on the corresponding free parameter, associated to the cosmological constant. On the other hand, in the unimodular case, the value of the cosmological constant, which will first enter the theory at the level of the equations of motion, is also a free parameter. However none of the predictable values of the irrelevant couplings depend on that parameter.
Furthermore, as pointed out in \cite{Eichhorn:2013xr}, choosing $\sqrt{g}= \epsilon$ as a gauge in quantum Einstein gravity implies that a corresponding Faddeev-Popov ghost contributes to the Renormalization Group flow. This additional ghost is completely absent in unimodular gravity. Imposing unimodularity directly on the allowed configurations of the metric in the path-integral also changes the symmetry from diffeomorphisms to transverse diffeomorphisms; again implying differences at the quantum level. Choosing a classically equivalent formulation of unimodularity can allow to keep full diffeomorphism invariance \cite{Henneaux:1989zc}, however we focus on the other case here.

Finally, imposing $\sqrt{g}= \epsilon$ in unimodular gravity also implies that the spectrum of quantum fluctuations, i.e., the off-shell part of the propagator of metric fluctuations, differs. We will again see this explicitly in Sec.~\ref{sec:secondvariation}. By itself, this already changes the RG flow.
We conclude that an equivalence between unimodular quantum gravity and quantum Einstein gravity is not be expected.

\subsection{Deriving the flow equations}
In this section, we will detail the derivation of the RG flow equation for unimodular quantum gravity. We will focus on a setting where unimodularity is implemented as a restriction on the allowed configurations in the path integral. This has the advantage that it reduces the number of fluctuating gauge degrees of freedom in the path integral, and could therefore be expected to yield better results already in simple approximations of the full path-integral. 

\subsubsection{Wetterich equation}
To examine whether asymptotic safety is realized in unimodular gravity, knowledge of the non- perturbative beta functions is required. Here, we will use the framework of the functional Renormalization Group to obtain these. In that setting, an infrared cutoff function $R_k(-\mathcal{D}^2)$, (with $\mathcal{D}$ denoting a placeholder for the appropriate covariant Laplacian) depending on the momentum scale $k$, is included in the generating functional, which suppresses quantum fluctuations with covariant momenta $-\mathcal{D}^2<k^2$ \cite{Wetterich:1993yh,Morris:1993qb}. For instance, a theta-cutoff takes the form \cite{Litim:2001up}
\be
R_k(-\mathcal{D}^2)= (k^2- (-\mathcal{D}^2)) \theta(k^2-(-\mathcal{D}^2)).
\ee
This allows us to obtain the scale dependent effective action $\Gamma_k$, which encodes the effect of high-momentum quantum fluctuations. Its scale dependence is governed by the Wetterich-equation
\be
\partial_t \Gamma_k = \frac{1}{2} {\rm STr} \left(\Gamma_k^{(2)}+R_k \right)^{-1} \partial_t R_k, \quad  \quad \partial_t=k \partial_k. 
\ee
Herein $\Gamma_k^{(2)}$ is the second functional derivative with respect to the quantum fields, which is matrix-valued in field space. The supertrace implies a summation in field space, including a negative sign for Grassmann-valued fields. It also encodes a summation/integration over the discrete/continuous eigenvalues of the kinetic operator $\Gamma_k^{(2)}$. 
For reviews, see, e.g.,  \cite{rg_reviews}. In the case of gravity, M.~Reuter has pioneered the application of the Wetterich equation in \cite{Reuter:1996cp}. Gravity-specific reviews of the asymptotic safety scenario and the application of the functional Renormalization Group can be found in, e.g., \cite{ASreviews}.

The application of Renormalization Group methods requires us to set a scale, which seems a challenging task in quantum gravity, where no background spacetime is assumed to exist. Here, the background field method \cite{Abbott:1980hw} can be employed to circumvent this problem: Splitting the full metric into a background and a fluctuation piece provides a background covariant derivative. This can be used to define ''high-momentum" and "low-momentum" quantum fluctuations by decomposing the quantum field into eigenfunctions of the background Laplacian and sorting them according to their eigenvalue. At the same time, admitting fluctuations of arbitrarily large amplitude implies that we can still perform the functional integral over all metric configurations, as long as the topology is kept fixed. While quantum Einstein gravity admits a linear split of the metric, unimodularity requires us to use a non-linear split of the form
\be
g_{\mu \nu} = \bar{g}_{\mu \kappa} {\rm exp}(h^._.)^{\kappa}_{\nu} = \bar{g}_{\mu \nu} + \bar{g}_{\mu \kappa}h^{\kappa}_{\nu} + \frac{1}{2}\bar{g}_{\mu \kappa} h^{\kappa \lambda}h_{\lambda\nu}+... = \bar{g}_{\mu \nu} + h_{\mu \nu} + \frac{1}{2} h_{\mu \kappa} h^{\kappa}_{\nu}+...\label{bckgrfluc},
\ee
where the background metric $\bar{g}_{\mu \nu}$ is used to lower and raise indices at each order of the expansion in the fluctuation field $h_{\mu \nu}.$
We then take the path-integral over the fluctuation field $h_{\mu \nu}$ as the \emph{definition} of the generating functional for quantum gravity.
(The effect of using such a decomposition in quantum Einstein gravity has been studied in in \cite{Nink:2014yya,Codello:2014wfa} and   \cite{Percacci:2015wwa}.)
Note that a similar decomposition has to be invoked for calculations involving fluctuations, e.g., in a cosmological setting, in the context of (semi-) classical unimodular gravity.

We then have at our disposal the background covariant Laplacian $-\bar{D}^2$ which we can use to set up a regulator $R^{\mu \nu \kappa \lambda}(-\bar{D}^2+\alpha \Rb$ (not to be confused with the Riemann tensor) for the fluctuation field
\be
h_{\mu \nu}R^{\mu \nu \kappa \lambda}(-\bar{D}^2+\alpha \Rb)h_{\kappa \lambda},
\ee
where $\alpha$ is the prefactor of a possible additional dependence on the background curvature $\Rb$. While the action is invariant under a simultaneous transformation of the background and fluctuation piece ($\bar{g}_{\mu \kappa} \rightarrow \bar{g}_{\mu \lambda}{\rm exp}(\gamma^._.)^{\lambda}_{\kappa}, \, h_{\mu \kappa} \rightarrow h_{\mu \kappa} - \gamma_{\mu \kappa}  +1/2 [h_{..},\gamma_{..}]_{\mu \kappa} -1/12 [h_{..},[h_{..},\gamma_{..}]]_{\mu \kappa}+1/6[\gamma_{..},[\gamma_{..},h_{..}]]_{\mu \kappa}+...$), the regulator term is clearly not. The same will be true for the gauge-fixing, as we will employ a background gauge fixing here. This implies that background-field couplings and fluctuation-field couplings will satisfy different flow equations. For instance, 
 the RG flow of the background Newton coupling will differ from that of the prefactor of the term quadratic in the fluctuation field and in derivatives. 
In this work, we will not resolve this difference, but instead identify background and fluctuation-field couplings, and leave the next step to future work. Thus, while a full calculation would feature beta functions for the background and fluctuation couplings where the nontrivial terms can only depend on the fluctuation couplings, our approximation will involve a nontrivial dependence on the background couplings.

\subsubsection{Second variation of the action}\label{sec:secondvariation}
Our truncation is given by
\be
\Gamma_k = \int d^4x \epsilon\, f(R).\label{truncation}
\ee
Within quantum Einstein gravity, an analogous truncation has been considered in \cite{Codello:2007bd, 
Machado:2007ea,Benedetti:2012dx,Dietz:2012ic,Falls:2013bv}. 

The unimodularity condition implies that the number of possible terms in the variation of \Eqref{truncation} will be reduced, since terms which are present in quantum Einstein gravity, such as $(\delta^2 \sqrt{g}) f(R)$ and $(\delta \sqrt{g})f'(R) \delta R$, do not exist here. In fact
\be
\delta^2 \Gamma_k = \int d^4x \epsilon \left( f'(R) \delta^2 R + f''(R) (\delta R)^2\right).
\ee
Evaluating the variation, starting from the relation \Eqref{bckgrfluc}, and using a $4$-sphere for the background field configuration\footnote{For this configuration, we have $\Rb_{\mu \nu} = \frac{\Rb}{4} \gb_{\mu \nu}$ and $\Rb_{\mu \nu \kappa \lambda} = \frac{\Rb}{12} \left( \gb_{\mu \kappa} \gb_{\nu \lambda} - \gb_{\mu \lambda}\gb_{\nu \kappa}\right)$.}, we obtain
\bea
\Gamma^{2}& =& \frac{1}{2} \int d^4 x \epsilon \, \Bigl[f''(R) h_{\mu \nu}\Db^{\mu}\Db^{\nu}\Db^{\kappa}\Db^{\lambda}h_{\kappa \lambda}+f'(R) \left(- \frac{1}{12}\Rb\, h_{\mu \nu}h^{\mu \nu}- h_{\mu \nu}\Db^{\mu}\Db^{\lambda}h_{\lambda}^{\nu} + \frac{1}{2}h_{\mu \nu}\Db^2 h^{\mu \nu} \right) \Bigr].
\eea
Herein, $g_{\mu \nu} = \bar{g}_{\mu \nu}$, i.e., we employ a single-metric approximation from now on.

As a next step, we insert a York-decomposition of the fluctuation field into a transverse traceless tensor, a transverse vector, and a scalar (corresponding to the longitudinal vector mode). Note that in contrast to the usual case, there is no trace mode, i.e.,
\be
h_{\mu \nu} = h_{\mu \nu}^{TT} + \Db_{\mu}v_{\nu}+ \Db_{\nu}v_{\mu}+ \Db_{\mu}\Db_{\nu} \sigma - \frac{1}{4} \gb_{\mu \nu}\Db^2 \sigma,
\ee
where $\Db^{\nu}h_{\mu \nu}^{TT}= 0$, $\gb^{\mu \nu}h_{\mu \nu}^{TT}=0$ and $\Db^{\mu}v_{\mu}=0.$

It turns out that the second variation evaluated on the transverse vector mode vanishes. In other words, the dynamics of the vector mode is arising from the gauge-fixing term only, i.e., it is ``pure gauge". This is another major difference to the case of $f(R)$ truncations in quantum Einstein gravity, see, e.g., \cite{Codello:2007bd, 
Machado:2007ea,Benedetti:2012dx}.

For the transverse traceless tensor mode we obtain
\be
\frac{1}{2}\int d^4x \epsilon\, h_{\mu \nu}\Gamma^{(2)\, \mu \nu \kappa \lambda}_{TT}h_{\kappa \lambda}= \frac{1}{2} \int d^4 x \epsilon f'(R) h_{\mu \nu}^{TT}\left(\frac{1}{2}\Db^2 -\Rb \frac{1}{12} \right)h^{TT\,\mu \nu}.
\ee

Finally, the scalar mode is governed by the following dynamics
\bea
\frac{1}{2}\int d^4x \epsilon\, \sigma\Gamma^{(2)}_{\sigma \sigma} \sigma &=&\frac{1}{2} \int d^4x \epsilon \sigma \Bigl[ f'(\Rb) \left(\frac{-1}{16} \Rb \Db^2 \Db^2 - \frac{3}{16} \Db^2 \Db^2 \Db^2 \right) \nonumber\\
&{}&+ f''(\Rb) \left(\frac{9}{16}\Db^2\Db^2\Db^2\Db^2 +\frac{3}{8}\Rb \Db^2 \Db^2 \Db^2 + \frac{1}{16}\Rb^2 \Db^2\Db^2 \right)
\Bigr] \sigma.
\eea

As usual, no mixed contributions $\Gamma_{\sigma v}$ etc. can exist because of the transversality and tracelessness of $h_{\mu \nu}^{TT}$ and $v_{\mu}$.

\subsubsection{Gauge-fixing}
We choose a gauge-fixing that is related to the harmonic gauge condition, but modified such that it satisfies
\be
\bar{g}^{\mu \nu}\bar{D}_{\nu}F_{\mu}=0,
\ee
for the spherical background. Accordingly, this choice of gauge fixing only imposes three instead of four gauge-conditions, i.e., it only fixes the transversal diffeomorphisms, infinitesimally defined by
 \begin{equation}
\delta_D g_{\mu \nu} = \mathcal{L}_v g_{\mu \nu} \mbox{ with } D_{\mu}v^{\mu}=0.\label{eq:transdiff}
\end{equation}
Note that in models of gravity which are invariant under transverse diffeomorphism, an additional scalar mode appears upon linearization. As noted in \cite{Alvarez:2006uu}, this mode is absent in two cases: If the symmetry is enhanced to  full diffeomorphism symmetry, yielding standard Einstein gravity, or if the metric determinant remains fixed. Then the additional scalar, which plays the role of the determinant, is removed from the model.

Gauge-fixing only the transverse diffeomorphisms is achieved by using the longitudinal and transversal projectors defined in \cite{Benedetti:2010nr}, which read
\bea
\Pi_{L\, \mu \nu} &=& - \Db_{\mu} \left( -\Db^2\right)^{-1}\Db_{\nu},\\
\Pi_{T\, \mu \nu} &=& \gb_{\mu \nu} - \Pi_{L\, \mu \nu}.
\eea
As they should, these satisfy $\Pi_{L\, \mu \nu} \Pi_{L\, \kappa}^{\,\,\nu} = \Pi_{L\, \mu \kappa}$, $\Pi_{L\, \mu \nu} \Pi_{T\, \kappa}^{\,\,\nu} =0$ and 
$\Pi_{T\, \mu \nu} \Pi_{T\, \kappa}^{\,\,\nu} = \Pi_{T\, \mu \kappa}$. 
We now project the harmonic gauge on its transversal part \cite{Alvarez:2008zw} and define
\be
F_{\mu} = \sqrt{2}\Pi_{T\, \mu}^{\,\, \kappa}\bar{D}^{\nu}h_{\nu \kappa}.
\ee
 It is then straightforward to see that $\bar{g}^{\mu \nu}\bar{D}_{\nu}F_{\mu}=0$. Accordingly only three conditions are imposed on the fluctuation field, which one can easily see by inserting the York decomposition: It turns out that the gauge fixing does \emph{not} impose a condition on $\sigma$, but only on $v_{\mu}$, which has only three independent components. These turn out to be gauge modes.
Indeed 
\be
F_{\mu} =\sqrt{2} \left(\bar{D}^2 + \frac{\bar{R}}{4} \right)v_{\mu}.
\ee
Thus the gauge-fixing action reads
\be
S_{gf\, v}= \frac{1}{\alpha} \int d^4x \epsilon v_{\mu} \left(\bar{D}^2+\frac{\Rb}{4} \right)^2 v^{\mu}.
\ee
Finally, the Faddeev-Popov ghost action is obtained in the usual way and reads
\be
S_{gh} = - \int d^4x \epsilon\, \bar{c}^{\mu} \left(\bar{D}^2 + \frac{\Rb}{4} \right)c_{\mu},
\ee
where we have already identified $g_{\mu \nu} = \bar{g}_{\mu \nu}$ and $\bar{D}_{\mu} c^{\mu} =0 = \bar{D}_{\mu} \bar{c}^{\mu}$. (As we only evaluate the ghost loop contribution to the running in the gravitational background couplings, this is already allowed at this stage.)

\subsubsection{Jacobian and auxiliary fields}
The York decomposition implies the existence of a nontrivial Jacobian in the generating functional \cite{Lauscher:2001ya}. Here, we will deal with this Jacobian by employing the following strategy: From the structure of $S_{gf\, v}$ it is obvious that a part of the Jacobian can be cancelled by the field redefinition
\be
v_{\mu} \rightarrow \sqrt{-\Db^2 - \frac{\Rb}{4}}v_{\mu}.
\ee
Employing this field redefinition results in $S_{gf\, v} =-1/{\alpha}\int d^4x \epsilon v_{\mu} \left(\bar{D}^2+\frac{\Rb}{4} \right) v^{\mu}.$, i.e., the vector mode does not contribute to the RG flow if we impose Landau gauge. In principle, we could choose to nevertheless impose a regulator on that mode with a dependence on the gauge parameter. Here, we take the point of view that a vanishing (unregularized) propagator allows us to trivially integrate out the $v$ mode in the path-integral, such that it does not affect the effective action. (Alternatively, a gauge-choice of the form $v_{\mu}=0$ could also be imposed, as in \cite{Percacci:2015wwa}, also leading to a vanishing contribution of the vector mode.)

On the other hand, a corresponding redefinition of $\sigma$ in order to absorb the remaining part of the Jacobian would not lead to a simple form of the inverse propagator. Accordingly we introduce auxiliary fields to take into account that part of the Jacobian. The corresponding action is given by
\bea
\Gamma_{k\, \rm aux} &=& \int d^4x \epsilon \Bigl[\frac{3}{4} \bar{\chi} \left(-\Db^2- \frac{\Rb}{3} \right) (-\Db^2) \chi +\frac{3}{4} \zeta \left(-\Db^2 - \frac{\Rb}{3} \right) (-\Db^2) \zeta  \Bigr],
\eea
where $\chi$ is a complex Grassmann field and $\zeta$ is a real scalar field. These give the same contribution to the flow equation, with a relative factor of $-2$. Accordingly, the total contribution comes with a factor $-1/2$.

\subsubsection{Choice of two regularization schemes and evaluation of traces}
We will study two different regulators in the following. 
For the first choice, we follow \cite{Benedetti:2012dx} and
employ regulators which essentially substitute the following covariant Laplace-type operators by $k^2$:
\bea
\Delta_2 &=& \Delta + \frac{2 \Rb}{12}, \quad \Delta_1 = \Delta - \frac{\Rb}{4}, \quad  \Delta_0 = \Delta - \frac{\Rb}{3}, \label{eq:LaplS}
\eea
for the transverse traceless tensor, the transverse vector, and the scalar. Herein $-\Db^2 = \Delta$.

We choose a Litim-type cutoff \cite{Litim:2001up}
\be
R_{k\, TT\, 1} = -\frac{1}{2}f'(R)(k^2- \Delta_2)\theta(k^2-\Delta_2)),
\ee
for the transverse traceless tensor.
Note that the negative sign is exactly as it should be, as in the simplest case $f(R) = \frac{-1}{16 \pi G}R$.
For the scalar mode we obtain the slightly lengthier expression
\bea
R_{k\, \sigma\, 1} &=&\Bigl[ f''(R)\left(\frac{R^2}{16}(k^4-\Delta_0^2) +\frac{3}{8} R (k^6-\Delta_0^3) +\frac{9}{16} (k^8-\Delta_0^4) \right)\nonumber\\
&{}& + f'(R) \left( \frac{1}{48} R^2 (k^2-\Delta_0) +\frac{1}{8} R (k^4-\Delta_0^2) +\frac{3}{16} (k^6-\Delta_0^3)\right)\Bigr] \theta(k^2-\Delta_0),
\eea
such that for $k^2 - \Delta_0>0$
\be
\Gamma^{(2)}_{k\, \sigma} +R_{k\, \sigma\,1} \rightarrow f''(R) \left(\frac{R^2\, k^4}{16}+\frac{3}{8} R\, k^6 +\frac{9}{16} k^8 \right) + f'(R) \left( \frac{1}{48} R^2\, k^2 +\frac{1}{8}R\, k^4 +\frac{3}{16} k^6\right).
\ee
Since the function $f(R)$ appears explicitly, its scale-derivatives will feature on the right-hand side of the Wetterich equation. They will lead to a schematic structure of the form $\partial_t g = c_1 g^2 + c_2 \partial_t g+\dots$ of the flow equation for the couplings. This results in nonperturbative resummation structures, i.e., $\partial_t g \sim \frac{c_1 g^2}{1-c_2}$. As the prefactor $c_2$ can contain further couplings, this choice of "spectrally adjusted" \cite{Litim:2002xm,Gies:2002af} regulator roughly corresponds to a resummation of an infinite series of polynomial terms in the couplings.

Explicitly, we will use the following derivatives
\bea
\partial_t f'(R) &=& k^2 \left(2 \tf'+\partial_t \tf' -2 \tR \tf'' \right),\\
\partial_t f''(R) &=&\partial_t \tf''-2\tR \tf'''.
\eea
in terms of the dimensionless function $\tf(\tR) = k^{-4} f(R)$, where $\tR = \frac{R}{k^2}$.

For the auxiliary fields and ghost we take
\bea
R_{k\,{\rm gh}\,1} &=& - \sqrt{2}(k^2-\Delta_1)\theta(k^2-\Delta_1),\\
R_{k\,{\rm aux}\,1}&=& \frac{3}{4} \left( k^4-\Delta_0^2 +\frac{R}{3} \left( k^2-\Delta_0\right)\right)\theta(k^2-\Delta_0).
\eea
In order to test the reliability of our results, we will actually perform a fixed-point search with two different regularization schemes. As our second choice we employ a regulator, which essentially substitutes covariant Laplacians $\Delta = -\Db^2$ with $k^2$ in the regularized propagator. The additional curvature dependence which is introduced in the regulator when using the operators \Eqref{eq:LaplS} is removed in this choice, resulting in a shift in possible poles of the flow equation. With $\Delta = - \bar{D}^2$, this choice corresponds to
\be
R_{k\, TT\, 2} = -\frac{1}{2}f'(R)(k^2- \Delta)\theta(k^2-\Delta)),
\ee
for the transverse traceless tensor.
For the scalar mode, we choose a regulator of the form
\bea
R_{k\, \sigma,\,2}&=& \Bigl[f'(R) \left(\frac{-1}{16} R (k^4-\Delta^2) + \frac{3}{16} (k^6-\Delta^3)\right)\nonumber\\
&{}& +f''(R) \left(\frac{9}{16}(k^8-\Delta^4) - \frac{3}{8} R (k^6-\Delta^3) + \frac{R^2}{16} (k^4-\Delta^2) \right)  \Bigr] \theta(k^2- \Delta).
\eea
For the ghost, we choose
\be
R_{k\, \rm gh,\, 2}= \sqrt{2}(-k^2+\Delta) \theta(k^2-\Delta).
\ee
The auxiliary fields come with a regulator of the form
\be
R_{k\, \rm aux,\, 2} = -\frac{3}{4} \left(-k^4+\Delta^2 +\frac{R}{3}(k^2- \Delta) \right)\theta(k^2-\Delta).
\ee

In both regularization schemes, we sum over the eigenvalues of the corresponding Laplacians, which can be obtained for both $\Delta$ and $\Delta_s$ from the following table, where the multiplicities are not affected by the curvature-dependent shift between $\Delta$ and $\Delta_s$.
\begin{table}[!here]
{\renewcommand{\arraystretch}{1.6}
\begin{tabular}{c|c|c}
& eigenvalue & multiplicity \\ \hline \hline
$\Delta_0$ & $\frac{n(n+3)-4}{12} R$; $n=0,1,...$& $\frac{(n+2)!(2n+3)}{6 n!}$ \\ \hline
$\Delta_1$ & $\frac{n(n+3)-4}{12}R$; $n=1,2,...$ & $\frac{(n+1)! n(n+3)(2n+3)}{2(n+1)!}$ \\ \hline
$\Delta_2$ & $\frac{n(n+3)}{12}R$; $n=2,3,...$ & $\frac{5(n+1)! (n+4)(n-1)(2n+3)}{6 (n+1)!}$\\ \hline
\end{tabular}
}
\caption{\label{eigtab} Eigenvalues and multiplicities of Laplacians acting on transverse traceless tensors, transverse vectors, and scalars, \cite{Rubin:1984tc}.}
\end{table}

 To convert the sum over eigenvalues into an integral, we employ the Euler-MacLaurin formula. In this step, additional terms which depend on the derivatives of the integrand at the lower boundary, arise. No contributions from the upper boundary exist, as $\theta(k^2-x) =0$ for $x \rightarrow \infty$. From the lower boundary, only the first few terms contribute: As $\partial_t R_k (\Gamma_k^{(2)}+R_k)$ is a polynomial of finite order in the eigenvalues, only the lowest few orders in derivatives, when evaluated at the lower boundary, can contribute.  

\section{Results: Asymptotic safety in $f(R)$}\label{sec:results}
\subsection{Flow equations}
The flow equation for $\tf(\tR)$ reads
\bea
\partial_t \tf(\tR)&=& -4 \tf(\tR)+2 \tR \tf'(\tR) + \frac{\tR^2}{384 \pi^2} \left(\mathcal{F}_{TT} +\mathcal{F}_{\sigma}+ \mathcal{F}_{\rm gh}+ \mathcal{F}_{\rm aux}\right), 
\eea
where the contributions from transverse traceless tensors $\mathcal{F}_{TT}$, scalars $\mathcal{F}_{\sigma}$, auxiliary fields $\mathcal{F}_{\rm aux}$ and Faddeev-Popov ghosts $\mathcal{F}_{\rm gh}$ depend on the choice of regulator, as discussed above.

For the first choice of cutoff, which is a function of $\Delta_s$, we obtain
\bea
\mathcal{F}_{TT}&=&\left( \frac{89}{18}+ \frac{60}{\tR^2} - \frac{40}{\tR}\right) + \frac{1}{2 \tf'} \left( \frac{89}{18}+ \frac{20}{\tR^2} - \frac{20}{\tR} - \frac{127\tR}{567}\right) \left(2 \tf' + \partial_t \tf' -2 \tf'' \tR \right) ,\label{floweq1TT}\\
\mathcal{F}_{\sigma}&=&\frac{1}{ \left(\frac{1}{8} \tf'' (3+\tR)^2 + \frac{1}{24}\tf' (3+\tR)^2\right)} \cdot \nonumber\\
&{}& \cdot\Bigl[
\left(-\frac{271}{90}+ \frac{12}{\tR^2}(1+\tR) \right) \cdot\left(\tf'\left(\frac{9}{8}+\frac{1}{2} \tR+\frac{1}{24}  \tR^2\right)   +\tf'' \left( \frac{9}{2}+ \frac{9}{4}\tR+\frac{\tR^2}{4}\right)\right) \nonumber\\
&{}&+ \left( 2 \tf' +\partial_t \tf' -2 \tf'' \tR\right)\left( -\frac{29}{48} + \frac{27}{20 \tR^2} + \frac{39}{16 \tR} - \tR - \frac{3}{16}\tR^2 +\frac{12161 \tR^3}{6842880}\right)\nonumber\\
&{}& +\left( \partial_t \tf'' - 2 \tf''' \tR\right) \left(-\frac{21}{16} + \frac{9}{2 \tR^2} + \frac{81}{10 \tR} - \frac{23 \tR}{8} - \frac{9 \tR^2}{16}  - \frac{439573 \tR^4}{622702080}\right) 
\Bigr], \label{floweq1sigma}\\
\mathcal{F}_{\rm gh}&=&-2 \left(\frac{109}{30}+ \frac{36+24 \tR}{\tR^2} \right), \label{floweq1gh}\\
\mathcal{F}_{\rm aux}&=&\frac{6+\tR}{90 \tR^2 (3+\tR)} \left(-1080 + \tR (-1080 +271 \tR) \right)\label{floweq1aux}.
\eea

As a main structural difference to quantum Einstein gravity with a full dynamical metric, including a conformal factor \cite{Codello:2007bd,Machado:2007ea,Benedetti:2012dx}, one should note that the function $\tf$ does not appear on the right-hand side of the flow equation, i.e., the cosmological constant is not part of the theory space.

For the second regulator, that imposes cutoffs on $\Delta$, we obtain
\bea
\mathcal{F}_{TT} &=& \frac{1}{4536 \tR^2 \tf' \left(1+\frac{ \tR}{6} \right)} \Bigl[ -252 \tf' \left( -1080+\tR (360+\tR)\right)\nonumber \\ 
&{}& + \left(2\tf' +\partial_t \tf' -2 \tf'' \tR \right) \left( 45360 \tR +\tR \left( -22680 +\tR (-126 +311 \tR)\right)\right) \Bigr]\label{floweq2TT}\\
\mathcal{F}_{\sigma}&=&\frac{1}{2 \left(\frac{3-\tR}{16}\tf' + \tf'' \frac{9-6\tR+\tR^2}{16} \right)}\cdot \nonumber\\
&{}& \cdot\Biggl[ 
\frac{1}{16}(\partial_t \tf''-2 \tR \tf''') \left( -\frac{631}{10}+ \frac{72}{\tR^2} - \frac{72}{5\tR} + \frac{551 \tR}{15}- \frac{511 \tR^2}{90} - \frac{55189 \tR^4}{38918880}\right) \nonumber\\
&{}&+ \frac{1}{47900160 \tR^2} \left( 2 \tf' + \partial_t \tf' - 2 \tf'' \tR \right) \left( 64665216 +8981280 \tR -58977072 \tR^2 + 16997904 \tR^3 + 3815 \tR^5\right) \nonumber\\
&{}&+ \left(\tf' \left(\frac{9}{8} - \frac{\tR}{4} \right)+\tf'' \left( \frac{9}{2} - \frac{9}{4}\tR + \frac{\tR^2}{4}\right) \right) \cdot \left(-\frac{511}{90}+ \frac{4(3+\tR)}{\tR^2} \right) 
\Biggr]\label{floweq2sigma}
\\
\mathcal{F}_{\rm gh}&=&-\frac{8}{4-\tR} \left( -\frac{7}{60}+ \frac{6(6+\tR)}{\tR^2}\right), \label{floweq2gh}\\
\mathcal{F}_{\rm aux}&=&\frac{-6+\tR}{90\tR^2 (-3+\tR)} \left(-1080 +\tR (-360+511 \tR) \right). \label{floweq2aux}
\eea

Comparing the two equations for $\partial_t \tf(\tR)$, we note that the different choice of regularization scheme mainly serves the purpose of changing the singularity structure of the equation, which can play a role in the search for global solutions \cite{Benedetti:2012dx}. Apart from that, the structure is similar in both cases, with differences only in the numerical prefactors of most terms.

\subsection{Fixed points}
 To search for fixed points, we expand $\tf(\tR)$ in a polynomial around vanishing curvature, where the cosmological constant is again absent,
\be
\tf(\tR) = \sum_{n=1}^{10}a_n \tR^n.
\ee
 Note that the Newton coupling is given by $G = -\frac{1}{16 \pi a_1}$. Thus a negative fixed-point value for $a_1$ translates into a positive microscopic Newton coupling. 
Using the first regularization choice \Eqref{floweq1TT} - \Eqref{floweq1aux}, we find a number of fixed points, where we list only the most stable one in Tab.~\ref{FP1}. Beyond $\tR^4$ it becomes rather cumbersome to evaluate all solutions of the fixed-point equations. We thus pick the most stable fixed point that exists in the four truncations up to $a_4$. At $\tR^5$ and beyond we perform a numerical search for the solution of the fixed-point equation in the vicinity of the fixed-point coordinates at lower truncation order, and no longer investigate all solutions.

\begin{table}[!here]
\begin{tabular}{c|c|c|c|c|c|c|c|c|c}
$a_{1\, \ast}$ & $a_{2\, \ast}$& $a_{3\, \ast}$ & $a_{4\, \ast}$ & $a_{5\, \ast}$ &$a_{6\, \ast}$& $a_{7\, \ast}$ & $a_{8\, \ast}$ & $a_{9\, \ast}$ &$a_{10\, \ast}$ \\ \hline
-0.0121 \\ \hline \hline
-0.0118 & 0.0031 \\ \hline \hline
-0.0117 & 0.0037 & 0.00025 \\ \hline \hline
-0.0113 & 0.0038 & 0.00017 & $-5.40 \cdot 10^{-5}$ \\ \hline \hline
-0.0112 & 0.0039 & 0.00020 & $-5.81 \cdot 10^{-5}$ & $6.25 \cdot 10^{-6}$ \\ \hline \hline
-0.0111 & 0.0039 & 0.00018 & $-6.68 \cdot 10^{-5}$ & $5.11 \cdot 10^{-6}$& $-2.59 \cdot 10^{-6}$ \\ \hline \hline
-0.0112 & 0.0039 & 0.00018 & $ -6.67\cdot 10^{-5}$ & $ 4.77\cdot 10^{-6}$& $-2.63 \cdot 10^{-6}$&$-1.09 \cdot 10^{-7}$ \\ \hline \hline
-0.0111 & 0.0039 & 0.00018 & $- 6.82\cdot 10^{-5}$ & $ 4.15\cdot 10^{-6}$& $-3.12 \cdot 10^{-6}$&$-2.34 \cdot 10^{-7}$&$ -1.89\cdot 10^{-7}$ \\ \hline \hline
-0.0111 & 0.0038 & 0.00017 & $-6.85\cdot 10^{-5}$ & $3.72 \cdot 10^{-6}$ & $-3.26 \cdot 10^{-6}$ & $-3.54 \cdot 10^{-7}$ & $-2.24 \cdot 10^{-7}$ & $-4.91 \cdot 10^{-8}$\\ \hline \hline
-0.0111 & 0.0038 & 0.00017 & $-6.90\cdot 10^{-5}$ & $3.37 \cdot 10^{-6}$ & $-3.44 \cdot 10^{-6}$ & $-4.37 \cdot 10^{-7}$ & $-2.85 \cdot 10^{-7}$ & $-6.96 \cdot 10^{-8}$ & $-2.62 \cdot 10^{-8}$\\ \hline \hline
\end{tabular}
\caption{\label{FP1}Fixed-point values for the dimensionless couplings $a_{n}$ in a polynomial expansion around vanishing curvature, at increasing truncation order.}
\end{table}

The fixed point at $n=1$ is related to that found in \cite{Eichhorn:2013xr}, which neglected the additional terms arising from the Euler MacLaurin formula.
We clearly observe that the fixed point at truncation order $n+1$ is a continuation of the fixed point at order $n$, as the coordinates of the fixed points lie reasonably close to each other. 
This stability under extensions of the truncation indicates that this is not an auxiliary fixed point. 
 Interestingly, the fixed-point values of the couplings $a_n, \, n\geq 5$ are four orders of magnitude smaller than the highest-order coupling. This suggests that an approximation of the fixed-point action employing only the first few terms could be reasonable, for instance if cosmological consequences of unimodular asymptotic safety are deduced from "RG improved" calculations.

\begin{table}[!top]
\begin{tabular}{c|c|c|c|c|c|c|c|c}
 $\theta_{1,2}$ & $\theta_3$ & $\theta_4$ & $\theta_5$& $\theta_6$& $\theta_7$ & $\theta_8$ & $\theta_9$& $\theta_{10}$\\
2.295 \\ \hline \hline 
2.122 $\pm$ i 1.232\\ \hline \hline
2.778 $\pm$ i 1.232 &  -1.233 \\ \hline \hline
2.832 $\pm$ i 0.781 & -1.113 & -3.111 \\ \hline \hline
2.912 $\pm$ i 0.687   & -1.172 & -3.415 & -5.235\\ \hline \hline
2.863 $\pm$ i 0.654  & -1.155 & -3.328 & -5.447 & -6.997 \\ \hline \hline
2.862 $\pm$ i 0.671  & -1.197 & -3.382 & -5.380 & -7.464 & -8.863 \\ \hline \hline
2.847 $\pm$ i 0.673  & -1.204 & -3.418 & -5.483 & -7.325 & -9.398 & -10.708 \\ \hline \hline
2.841$\pm$ i 0.678 & -1.221 & -3.440 & -5.530 & -7.534 & -9.236 & -11.324 & -12.698\\ \hline \hline 
2.834 $\pm$ i 0.681  & -1.228 & -3.462 & -5.569 & -7.585 & -9.545 & -11.093 & -13.224 & -14.514 \\ \hline \hline
\end{tabular}
\caption{\label{theta1} Critical exponents at the fixed point shown in Tab.~\ref{FP1}. The first line is of course understood to feature only one real critical exponent.}
\end{table}

\begin{figure}[!top]
\includegraphics[width=0.48\linewidth]{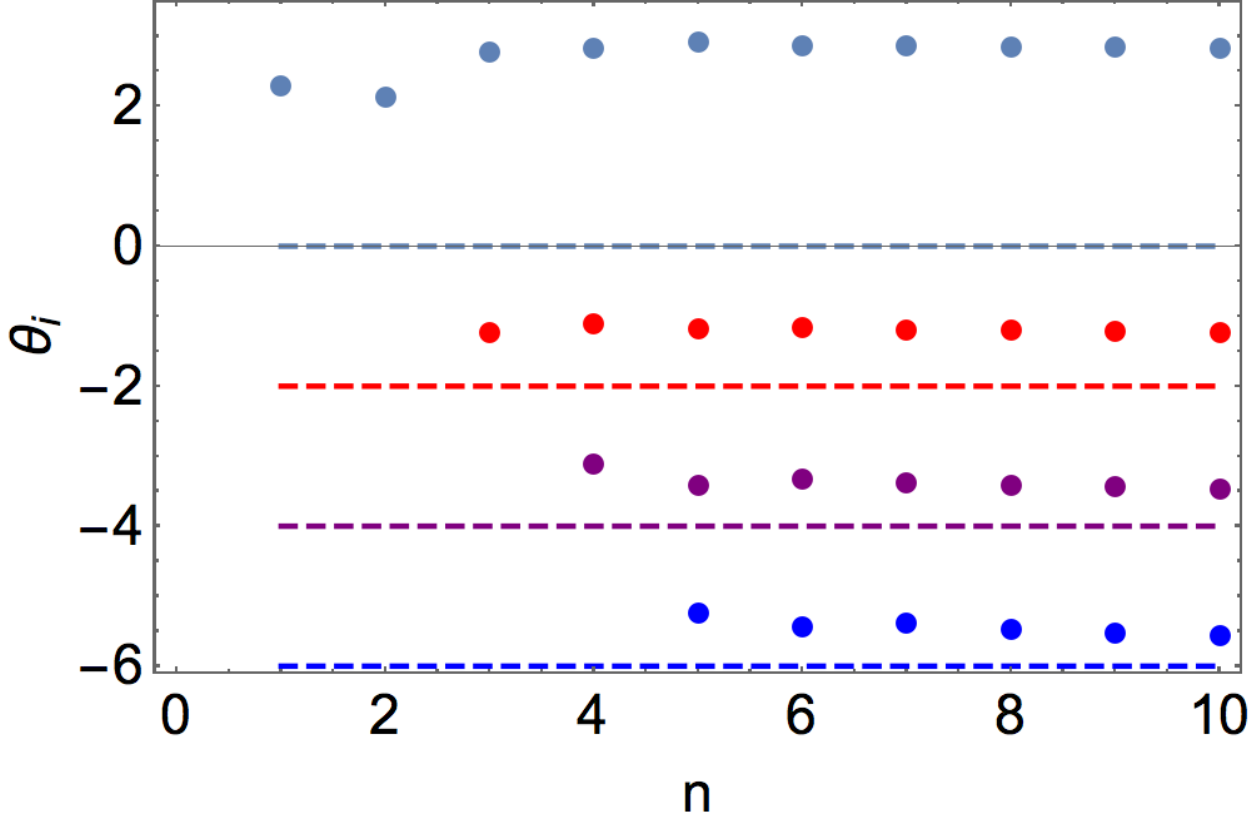} \includegraphics[width=0.48\linewidth]{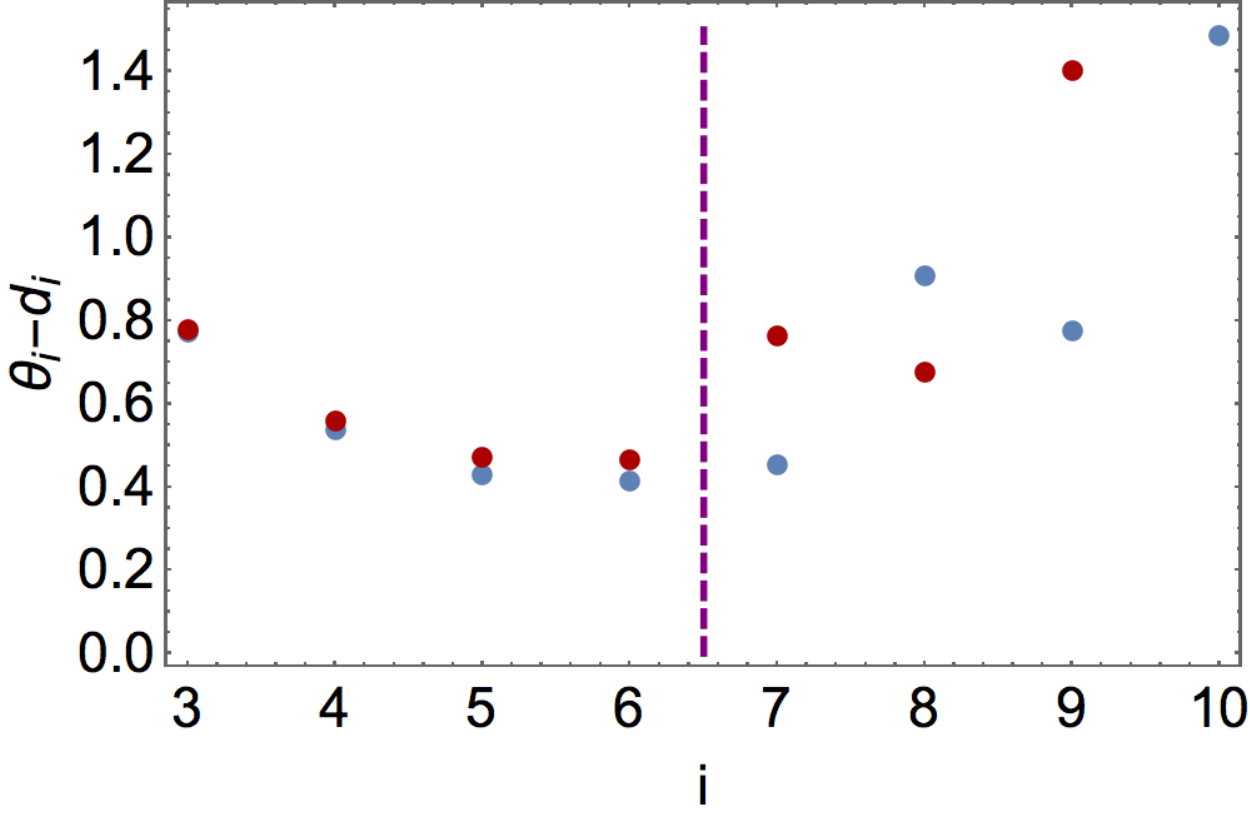}
\caption{\label{thetaplot1} Left panel: Real part of the five largest negative criitical exponents (the first two are degenerate) and  canonical dimensions of the couplings (dashed lines). Right panel: Difference between the critical exponents $\theta_i$ (starting with the third-largest) to the canonical dimension $d_i= -(2i-4)$ for the truncation including up to $\tR^9$ (dark red dots) and up to $\tR^{10}$ (blue dots). Beyond the dashed purple line our results have not converged.}
\end{figure}

The corresponding critical exponents are given in Tab.~\ref{theta1}, and again seem reasonably stable under extensions of the truncation, cf.~Fig.~\ref{thetaplot1}.
Similarly to the case of quantum Einstein gravity, two exponents with positive real part are observed, i.e., $R$ and $R^2$ form relevant directions. In this approximation, the two critical exponents with positive real part form a complex conjugate pair, which could hint at a necessary extension of the truncation to obtain better numerical precision. Higher-order operators become increasingly irrelevant. The anomalous dimension is a positive contribution to the critical exponents for all operators, i.e., quantum fluctuations shift all operators towards relevance.
Comparing the value of the critical exponents $\theta_i$ beyond $i=3$ to the canonical dimension $d_i=-(2i-4)$, we observe a decreasing distance  up to $\theta_6$, cf.~Fig.~\ref{thetaplot1}. Beyond, our truncation is not large enough to produce convergent results, as can be inferred from the comparison between the largest and second-largest truncation. Tentatively extrapolating this trend would indicate that canonical scaling could be recovered at large powers of the curvature.

We observe an interesting difference to quantum Einstein gravity, where the gap between the smallest relevant and largest irrelevant eigenvalue is $\Delta_{UV} = 5.58$ \cite{Falls:2013bv}. In our case, we find $\Delta_{UV\, \rm unimod} \approx 4.06$, i.e., a significantly reduced gap. This is related to the fact that canonically irrelevant operators are shifted further into irrelevance in quantum Einstein gravity, whereas they are shifted into relevance in unimodular asymptotic safety.

At this stage, we should comment on the approximations we have used.
There are several sources of uncertainty in our results: Firstly, using a spherical background to project on the function $f(R)$ entails ambiguities. Operators with the same number of derivatives are projected onto simultaneously:
On a sphere, derivatives of the curvature tensor and its contractions vanish. Further, the Weyl tensor vanishes, allowing to re-express the Riemann tensor in terms of the Ricci tensor and the Ricci scalar. Thus, the remaining ambiguity lies in the fact that on a 4-sphere $R_{\mu \nu} =\frac{R}{4} g_{\mu \nu}$. Accordingly we derive the spectrum of fluctuations, i.e., $\Gamma^{(2)}$, from an action of the form $f(R)$, without contributions from other tensor structures. Using a sphere to evaluate the traces and project onto powers of $R$ implies that we project on $R^2 + c_1 R_{\mu\nu}R^{\mu \nu}$ at fourth order in derivatives, and so on, with unknown coefficients $c_i$. This ambiguity is inherent in the setup of our calculation and is owing to a compromise between uniqueness of the projection and computability. Here, our calculation is on a par with similar challenges encountered by $f(R)$ approximations in quantum Einstein gravity.

To address this ambiguity, we make a useful observation: Had we considered beta-functions for the disentangled system which distinguished between $R^2$ and $R_{\mu\nu}R^{\mu \nu}$, etc., a fixed point in that system necessitates the existence of a fixed point in our simplified system. Furthermore, a relevant direction in the simplified system can only occur, if (at least) one relevant direction exists in the disentangled system. Thus, our estimate of the number of relevant directions provides a lower bound on the number of relevant directions in the disentangled system, and a discovery of a fixed point in our approximation is a necessary condition for the existence of a fixed point in the disentangled system.

Secondly, we have truncated the theory space very severly. This is important in two places: Firstly, our truncation is not closed in the sense that operators beyond those present in the truncation are generated on the right-hand-side of the Wetterich equation. Secondly, further operators also change the spectrum of fluctuations, i.e.,  $\Gamma^{(2)}$, which enters the evaluation of the beta-functions for the coefficients $a_i$. 

Thirdly, we have employed a so-called single-metric approximation. This consists in neglecting the fact that no background coupling should appear on the right-hand side of the Wetterich equation, as the functional derivatives necessary to obtain $\Gamma^{(2)}$ are taken with respect to the fluctuation field. Thus, fluctuation field couplings should appear on the right-hand side of the Wetterich equation. Performing the necessary evaluation of fluctuation-field and background-field flows at the level of an $f(R)$ truncation is beyond the scope of this paper.

All these approximations can affect the stability of our results. One could expect that if the errors introduced into the results by the approximations are small, then the regulator-dependence should not be too large.
In this spirit we study whether a fixed point with similar properties can also be found using the second regularization scheme, \Eqref{floweq2TT}-\Eqref{floweq2aux}. We find a fixed point with similar properties (see Tab.~\ref{FP2} and \ref{theta2}), thus strengthening the conclusion that the fixed point is not a truncation artifact. 

\begin{table}[!top]
\begin{tabular}{c|c|c|c|c|c|c|c|c|c}
$a_{1\, \ast}$ & $a_{2\, \ast}$& $a_{3\, \ast}$ & $a_{4\, \ast}$ & $a_{5\, \ast}$ &$a_{6\, \ast}$& $a_{7\, \ast}$ & $a_{8\, \ast}$ & $a_{9\, \ast}$ &$a_{10\, \ast}$ \\
-0.0087\\ \hline \hline
-0.0122 & 0.0020 \\
 \hline \hline
-0.0082 & 0.0026 & 0.00031 \\ \hline \hline
-0.0083 & 0.0005 & 0.00045 & $6.63 \cdot 10^{-5}$\\ \hline \hline
-0.0083 & 0.00048 & 0.00044 & $6.56 \cdot 10^{-5}$&$-5.61 \cdot 10^{-7}$ \\ \hline \hline
-0.0083 & 0.00044 & 0.00043 & $6.32 \cdot 10^{-5}$&$-5.27 \cdot 10^{-7}$&$-5.61 \cdot 10^{-7}$  \\ \hline \hline
-0.0083 & 0.00046 & 0.00044 & $6.39 \cdot 10^{-5}$&$-3.93 \cdot 10^{-7}$&$-4.60 \cdot 10^{-7}$ &$1.26 \cdot 10^{-7}$  \\ \hline \hline
-0.0083 & 0.00047 & 0.00044 & $6.42 \cdot 10^{-5}$&$-3.901\cdot 10^{-7}$&$-3.84 \cdot 10^{-7}$ &$1.48 \cdot 10^{-7}$&$2.31 \cdot 10^{-8}$  \\ \hline \hline
-0.0083 & 0.00046 & 0.00044 & $6.41 \cdot 10^{-5}$&$-3.327\cdot 10^{-7}$&$-4.08 \cdot 10^{-7}$ &$1.30 \cdot 10^{-7}$&$1.90 \cdot 10^{-8}$ &$-5.23 \cdot 10^{-9}$  \\ \hline \hline
-0.0083 & 0.00046 & 0.00044 & $6.40 \cdot 10^{-5}$&$-3.51\cdot 10^{-7}$&$-4.32 \cdot 10^{-7}$ &$1.20 \cdot 10^{-7}$&$1.26 \cdot 10^{-8}$ &$-6.93 \cdot 10^{-9}$ &$-2.19 \cdot 10^{-9}$ \\ \hline \hline
\end{tabular}
\caption{\label{FP2} Fixed-point values using the second regularization scheme.}
\end{table}

\begin{table}[!top]
\begin{tabular}{c|c|c|c|c|c|c|c|c|c}
 $\theta_1$ & $\theta_2$ & $\theta_3$ & $\theta_4$ & $\theta_5$& $\theta_6$& $\theta_7$ & $\theta_8$ & $\theta_9$& $\theta_{10}$\\
2.355 \\ \hline \hline
3.034 & 3.034 \\
 \hline \hline
2.441 & 11.349 & -1.740 \\ \hline \hline
4.683 & 1.498 & -1.952 & -3.444 \\ \hline \hline
4.542 & 1.353 & -2.099 & -3.611 & -5.329 \\ \hline \hline
3.937 & 1.006 & -2.460 & -3.708 & -5.076 & -7.156 \\ \hline \hline
3.899 & 1.204 & -2.646 & -3.656 & -5.209 & -6.923 & -9.058 \\ \hline \hline
3.969 & 1.296 & -2.653 & -3.642 & -5.414 & -7.051 & -8.854 & -10.999 \\ \hline \hline
3.953 & 1.293 & -2.711 & -3.647 & -5.388 & -7.224 & -8.936 & -10.791 & -12.947 \\ \hline \hline
3.933 & 1.289 & -2.762 & -3.653 & -5.385 & -7.192 & -9.089 & -10.851 & -12.746 & -15.907 \\ \hline \hline
\end{tabular}
\caption{\label{theta2}Critical exponents at the fixed point shown in Tab.~\ref{FP2}.}
\end{table}

The fixed point found at positive Newton coupling with the second regularization scheme shares several characteristics with the fixed point found in the other scheme. Firstly, there are two positive critical exponents which are comparable to the real part of the two relevant critical exponents in the other scheme. We observe that fixed-point values and critical exponents converge slower than using the first regularization scheme, which could suggest that the first scheme is actually better adapted to the properties of unimodular $f(R)$ gravity.
Secondly, the fixed-point values for the couplings $a_n$, $n>4$, are at least four orders of magnitude smaller than the leading fixed-point values. Comparing the fixed-point coordinates in the two regularization schemes, one should keep in mind that all but the $a_2$ coupling have nonvanishing canonical mass-dimensionalities, i.e., the fixed-point values of all couplings should not be expected to be universal. 

With only two relevant directions, the number of free parameters is significantly smaller than that implied by the analysis in \cite{Saltas:2014cta}, with a considerable smaller truncation, where unimodularity was implemented together will full diffeomorphism symmetry, necessitating the introduction of Stueckelberg fields. This implies a significantly higher predictive power of our implementation, where $\sqrt{g} = \rm const$ is implemented as a restriction on the path-integral measure. It also clearly shows that while the two implementations are equivalent classically, one should not expect them to be so on the quantum level, see also \cite{Smolin:2009ti}. 

It is now interesting to examine the system while dropping the ''RG-improvement" terms $~ \partial_t \tf$ on the right-hand side, which arise as we chose a regulator that depends on $\tf$. Neglecting these terms can have no effect on the positions of the fixed points, but can change the value of the critical exponents. We observe that the signs of the critical exponents are stable -- another indication for the robustness of our result -- but their numerical values can change quite a bit. This is actually similar to the results in quantum Einstein gravity: Using the flow equations derived in \cite{Codello:2007bd,Benedetti:2012dx}, but dropping the RG improvement terms can produce different values for the critical exponents.

We conclude that the anomalous dimensions, related to $\partial_t \tf$, are rather large. This already suggests that it will be important to examine the system without employing a single-metric approximation, as this will allow to disentangle the background-beta functions from the anomalous dimensions, which are related to the fluctuation field\footnote{For a bimetric study of the gravitational Renormalization Group flow that uses the decomposition \Eqref{bckgrfluc} that is also required for the unimodular case, see \cite{Nink:2014yya}.}. As discussed in \cite{Bridle:2013sra}, a resolution of the difference between background field and fluctuation field could also play a major role in obtaining viable global solutions.

Our work also has implications for the asymptotic safety scenario in full gravity: It clarifies that the existence of a fixed point with three relevant directions in $f(R)$ truncations is not due to the dynamics of the conformal mode. While we in fact find one relevant direction less in unimodular gravity, this is due to the fact that we work in a different theory space, arising from a different definition of the degrees of freedom and the symmetries. In fact, simply dropping the contribution from the conformal mode in full gravity will not result in the removal of the relevant direction resulting from the presence of the cosmological constant.  It is the definition of a different theory space which does not contain the cosmological constant, which lowers the number of relevant directions.
Further, our result can be read as a further confirmation of asymptotic safety in quantum Einstein gravity: If the removal of only the conformal mode could destabilize the existence of a fixed point, one might conclude that the fixed point in quantum Einstein gravity was just a truncation artifact.

\subsection{Global solutions}
To analyze whether the fixed-point equation for $\tf(\tR)$ might admit global solutions, we bring it into normal form, i.e., we rewrite it such that the highest derivative, $\tf'''$, occurs with a prefactor normalized to one. 
It is a third-order differential equation, and thus a solution comes with three free parameters. Singularities in the equation decrease the number of parameters, as they impose constraints on the solution, if it is to be continued through the singularity. 
Let us first consider the second choice of regularization scheme, imposing a cutoff on $\Delta$. In that case, inspection of \Eqref{floweq2TT}- \Eqref{floweq2aux} reveals singularities at $\tR=0$, $\tR=4$ and $\tR=-6$. Additional singularities arise from the prefactor of the term $\sim \tf'''$, which has real zeros at $\tR \approx -0.916$ and $\tR \approx 1.624$. A seeming singularity at $\tR=3$ is actually not there, as it also occurs in the denominator of the prefactor of the term $\sim \tf'''$. So it does not in fact show up when we bring the equation into normal form. Note that while the equation for $\tf(\tR)$ was derived using a spherical background, we postulate its validity for $\tR<0$ here, and include singularities at negative $\tR$ in the parameter counting.
Taken together, these singularities already overconstrain the system, such that a global solution is unlikely to exist.
This situation is very similar to that found in \cite{Dietz:2012ic} analyzing the equation derived in \cite{Codello:2007bd}. As we have essentially followed a very similar procedure here, and in particular imposed the cutoff in a similar way, it is not unexpected that we find a differential equation with similar behavior. Following the insight in \cite{Benedetti:2012dx}, we can derive an alternative equation by adapting the choice of regulator, and imposing a cutoff on combinations such as $\Delta +\frac{R}{4}$, instead of on $\Delta$, which in fact corresponds to our first choice of regularization scheme.
In that case, we should consider \Eqref{floweq1TT}-\Eqref{floweq1aux} to determine the existence of fixed singularities. We again observe a fixed singularity at $\tR=0$. Additional singularities again arise from the prefactor of the $\tf'''$ term,  and lie at $\tR \approx -0.56$ and $\tR\approx 1.55$. 
At a first glance, another singularity seems to lie at $\tR=-3$, but again does not exist in the normal form of the equation.
We conclude that this third-order equation exhibits precisely three singularities, which suggests that a global solution could exist. Whether the asymptotic structure at large $\tR$ imposes further constraints, requires a more detailed analysis which we leave to the future. Here, we only observe that an asymptotic solution is given by $\tf(\tR) \sim A \tR^2$ at leading order, as expected from canonical dimensionality.

\section{Toward the real world: Adding matter}\label{sec:mattermatters}
\begin{figure}[!top]
\includegraphics[width=0.5\linewidth]{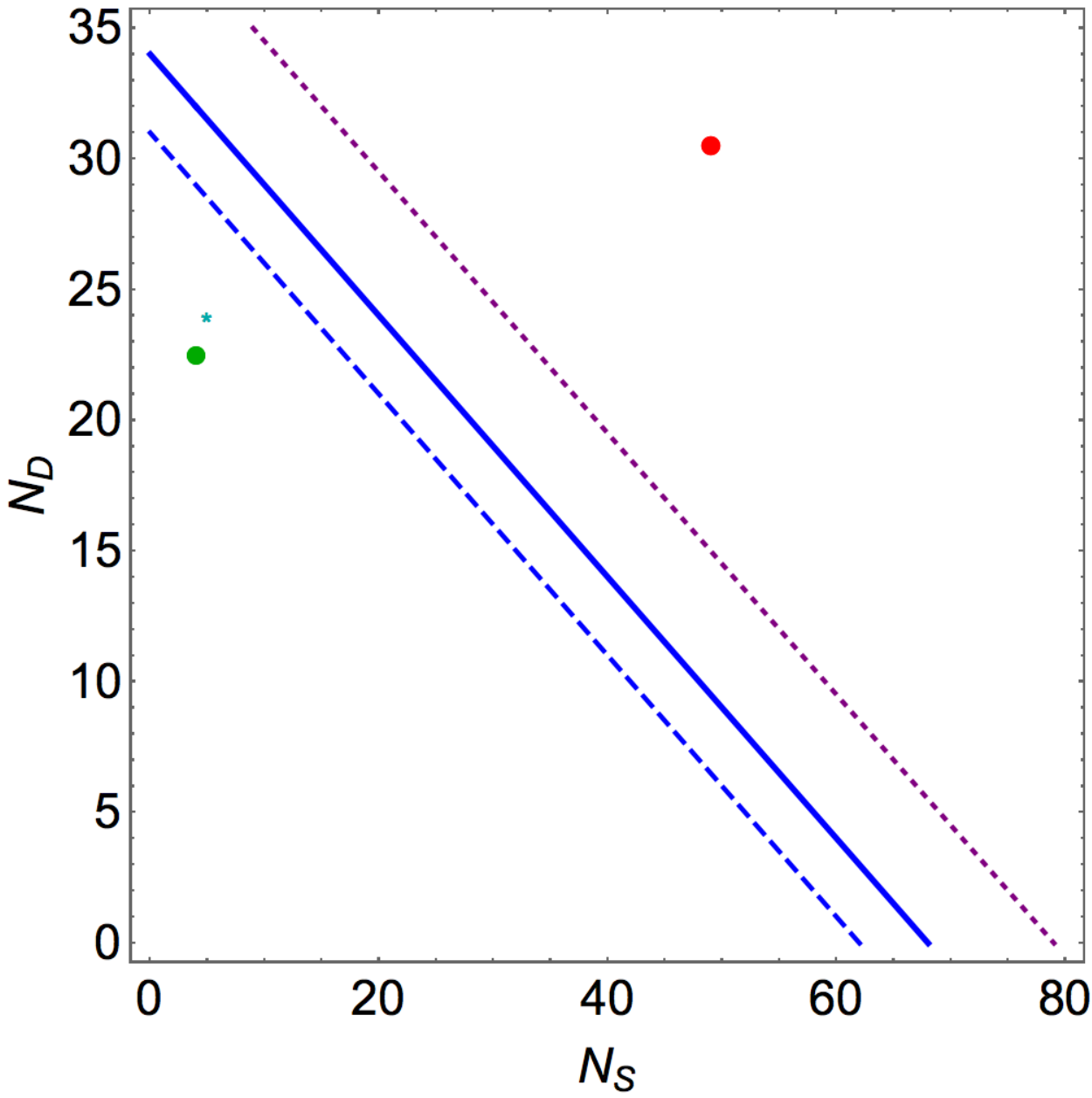}
\caption{\label{Matterplot} We show the bound on the allowed number of Dirac fermions $N_D$ and scalars $N_S$ with $N_V=12$ and $N_{RS}=0$. The blue thick line is obtained with the first regulator, and corresponds to \Eqref{eq:matterbeta}, while the thin dashed blue line corresponds to the second regulator. For comparison, the purple dashed line is the bound in quantum Einstein gravity from \cite{Dona:2013qba}. The green dot corresponds to the Standard Model degrees of freedom, and the turqoise asterisk additionally contains three right-handed fermions (neutrinos) and a dark matter scalar. The red dot, corresponding to the degrees of freedom of the MSSM lies outside the allowed region.}
\end{figure}
If a theory of quantum gravity is to be applied to our universe, dynamical matter degrees of freedom must be accounted for. Here we take the point of view that matter fields must be included at the microscopic level. Thus, our truncation should not only include dynamics for gravity, but also fermions, vector bosons, and scalars. We will consider only minimally coupled matter, and disregard further interactions for the moment. This is an approximation, as metric fluctuations induce matter self interactions in the UV \cite{Eichhorn:2011pc,Eichhorn:2012va}. Interestingly we can take advantage of  structural similarities between quantum Einstein gravity and unimodular gravity: The matter loop contribution to the background Newton coupling beta function is the same in the two settings. This will change within more sophisticated truncations. For instance, it is no longer true when wave-function renormalizations for the matter fields are included, as these are derived from different graviton-matter-vertices. We can thus take the matter contributions from \cite{Dona:2013qba, Dona:2014pla} and add these to the beta function for the background Newton coupling, where we use the $n=1$ truncation and identify $a_1=-1/(16 \pi G)$ and subsequently expand to second order in $G$.
We then have that
\be
\beta_G = 2 G - \frac{G^2}{6\pi} \left( 20-2N_D -N_S +4 N_V +N_{RS}\right),\label{eq:matterbeta}
\ee
where $N_V$ is the number of abelian vector fields, $N_D$ the number of Dirac fields, $N_S$ the number of real scalars and $N_{RS}$ the number of Rarita-Schwinger fields with spin 3/2.
Herein a type-II cutoff (nomenclature as in \cite{Codello:2007bd}) has been used for the matter fields, as it is required for the proper treatment of fermions \cite{Dona:2012am}.
For the gravitational contribution, we have used the cutoff imposed on $\Delta_S$. In analogy to the Einstein gravity case, there is a bound on the number of Dirac fermions and scalars, if the number of vectors and spin 3/2 fields is fixed. 
Disregarding Rarita-Schwinger fields,
\be
2N_D + N_S < 20 + 4N_V,\label{matterbound}
\ee
is required for a positive Newton coupling at the fixed point. While the microscopic value of this coupling is not (yet) restricted observationally, its infrared value must be positive. In all known truncations, the sign of the Newton coupling is preserved under the RG flow. We therefore exclude negative fixed-point values based on the positivity of the observed Newton coupling in the infrared.

Considering supersymmetric models, pure supergravity with a gravitino admits a viable fixed point. On the other hand, the matter content of the minimal supersymmetric standard model ($N_S=49$, $N_D=61/2$ and $N_V=12$) seems excluded. Whether this result will be affected when the truncation includes supersymmetric interactions, and also an appropriate regularization scheme \cite{Synatschke:2008pv} is used, is presently unclear.

Most importantly, the Standard Model matter content ($N_V=12$, $N_S=4$, $N_D= 45/2$) admits a gravitational fixed point at $G_{\ast}>0$. We can also add several additional scalars or fermions, that could constitute dark matter and still find a viable fixed point. This result is confirmed if we use a regulator imposed on $\Delta= -\bar{D}^2$ for the gravitational contribution. This changes the factor $20$ in \Eqref{eq:matterbeta} to $14$ and yields a tighter bound on fermions and scalars. Crucially, the bound still includes the Standard Model, cf.~Fig.~\ref{Matterplot}.
This indicates that unimodular quantum gravity can be compatible with the matter content of our universe.

While this result is obtained within the simplest truncation, and requires much further investigation, it is encouraging that we cannot find a way to exclude our model at this stage from phenomenological considerations of matter. On the other hand, we actually observe a tighter bound on matter than within quantum Einstein gravity. This arises, as the bound comes from a balance of gravitational and vector with fermion and scalar fluctuations. As the gravitational contribution is smaller in the unimodular case, the balance is reached for lower numbers of fermions/scalars.
If this result persists beyond our truncation, future discoveries of additional (dark matter) fermions and scalars might rule out unimodular asymptotic safety (or necessitate the introduction of additional vectors with the corresponding scalar modes to make them massive). 
Of course, much more detailed studies are necessary in order to investigate the reliability of these bounds, and also understand inhowfar these are universal. Crucially, the form of the bound presented in this paper should be understood as coming with considerable systematic errors.
Beyond its implications for unimodular quantum gravity, the bound \Eqref{matterbound} nevertheless points towards a potential option how experimental constraints might be put on quantum gravity models from low energy observations.

\section{Conclusions}\label{sec:conclusions}
We have studied the Renormalization Group flow of unimodular quantum gravity and found an interacting fixed point, further strengthening the evidence for unimodular asymptotic safety. 

In particular, we have used a truncation of the effective action to a function of the curvature scalar, $f(R)$, and derived the flow equation for this function. We have subsequently expanded in powers of the curvature up to $R^{10}$ and found a fixed point with two relevant directions, i.e., two out of ten couplings correspond to free parameters of the model. 

We have studied two different regularization schemes, and found a fixed point with similar properties and two relevant directions in both schemes. While this provides evidence that the fixed point is not an artifact of the approximation scheme, the remaining scheme dependence indicates that further studies are necessary to determine the fixed-point properties with quantitative precision. 

As a main difference to quantum Einstein gravity, the cosmological constant is not part of the theory space in the unimodular setting. The corresponding fine-tuning problem is therefore avoided. This makes unimodular asymptotic safety attractive from a phenomenological point of view. The cosmological constant will only enter, once the RG flow has been integrated into the infrared, where the effective equations of motion can be derived. As we have shown, these can then be reformulated with the help of the Bianchi-identities, and the cosmological constant appears as a constant of integration.

Furthermore the unimodularity condition $\sqrt{g} = \rm const$ reduces the number of propagating components of the metric in the quantum theory and most importantly removes the conformal mode from the path integral. This avoids the corresponding instability of the Euclidean path integral in an Einstein-Hilbert truncation. Besides, the reduction in degrees of freedom (by which we here mean components of the quantum field and not necessarily physical degrees of freedom) results in computational simplifications. Most importantly, we find that an appropriate gauge fixing removes all but the transverse traceless and one scalar mode of the metric. 

Performing a first step towards phenomenology, we have also studied the compatibility of unimodular asymptotic safety with minimally coupled matter degrees of freedom. Here we make use of structural similarities between quantum Einstein gravity and unimodular gravity at the level of the background couplings, which imply that the matter contributions to the running of the background Newton coupling in unimodular gravity agree with \cite{Dona:2013qba, Dona:2014pla}.
We find bounds on the number of allowed scalars and fermions, at a fixed number of Abelian vector bosons. At this order of the approximation it is only the gravitational contribution to the Renormalization Group flow that differs from the result in quantum Einstein gravity, and which yields a slightly tighter bound on the allowed number of matter fields. Most importantly, we find that within this approximation, the Standard Model, as well as small extensions by, e.g., a dark matter scalar, can be accommodated in unimodular asymptotic safety.
This suggests that unimodular asymptotic safety could pass an important observational test. Moreover, this provides a first example, where different models of quantum gravity might be distinguished observationally: As -- within simple truncations -- unimodular asymptotic safety and quantum Einstein gravity impose different bounds on the number of matter fields, detection of certain BSM-matter models, e.g., at the LHC, could potentially rule out one of these models while admitting the other. To further quantify this exciting possibility, nonminimal matter-gravity couplings need to be included in the truncations.

From here, several interesting routes are open toward future work: As has recently been emphasized in \cite{Nink:2014yya, Percacci:2015wwa}, the exponential parameterization of the metric in terms of the fluctuation field, that we employed here, can also be advantageous in quantum Einstein gravity. From our result, it is only a small step towards an analysis of the $f(R)$ truncation in that parameterization, as only the conformal mode needs to be added and the contribution from fluctuations of $\sqrt{g}$ needs to be added to the other propagators of the other modes.

Further, an extension of the truncation in a bimetric direction is possible as in quantum Einstein gravity \cite{bimetric} and seems indicated, see, e.g., \cite{Bridle:2013sra}. This is possible both within pure gravity as well as including matter. In that case, it is particularly interesting to investigate whether the unimodularity condition on the matter-graviton vertices can have observable consequences. For instance, one could imagine that differences in the vertices result in a change of the number of relevant directions in the gravity-matter sector. This could even lead to differences in predictions for the low-energy theory, e.g., along the lines of \cite{Shaposhnikov:2009pv}.

\emph{Acknowledgements} 
I would like to thank Rafael Sorkin, Robert Brandenberger, Christof Wetterich as well as Holger Gies and Tim Morris for useful discussions and Dario Benedetti for useful correspondence on \cite{Benedetti:2012dx}.
This work was supported by an Imperial College Junior Research Fellowship.

\end{document}